\documentclass[onecolumn,prb]{revtex4}
\usepackage{amsmath}
\usepackage{amsfonts}
\usepackage{amssymb}
\usepackage{graphics}
\usepackage{float}
\usepackage{graphicx}
\usepackage{epstopdf}
\usepackage[compact]{titlesec}
\usepackage{natbib}
\usepackage{threeparttable}
\usepackage[titletoc,title]{appendix}
\usepackage{lipsum}
\begin{document}
\title{A Probability-Density Function Approach to
Capture the Stochastic Dynamics of the
Nanomagnet and Impact on Circuit Performance}
\author{Nickvash Kani$^{1}$}
\email{nkani3@gatech.edu}
\author{Shaloo Rakheja$^{2}$}
\email{shaloo.rakheja@nyu.edu}
\author{Azad Naeemi$^1$}
\email{azad@gatech.edu}
\affiliation{$^{1}$ Electrical and Computer Engineering, Georgia Institute of Technology, Atlanta, GA, 30318 \\
$^{2}$ Electrical and Computer Engineering, New York University, New York, NY, 11201}

\begin{abstract}
In this paper we systematically evaluate the variation in the reversal delay of a nanomagnet driven by a longitudinal spin current while under the influence of thermal noise. We then use the results to evaluate the performance of an All-Spin-Logic (ASL) circuit. First, we review and expand on the physics of previously-published analytical models on stochastic nanomagnet switching. The limits of previously established models are defined and it is shown that these models are valid for nanomagnet reversal times $<$ 200 ps. Second, the insight obtained from previous models allows us to represent the probability density function (PDF) of the nanomagnet switching delay using the double exponential function of the Fr\'{e}chet distribution.
The PDF of a single nanomagnet is extended to more complex nanomagnet circuit configurations.
It is shown that the delay-variation penalty incurred by nanomagnets arranged in parallel configuration is dwarfed by the average delay increase for nanomagnets arranged in a series configuration. Finally, we demonstrate the impact of device-level performance variation on the circuit behavior using ASL logic gates. While the analysis presented in this paper uses an ASL-AND gate as the prototype switching circuit in the spin domain, the physical concepts are generic and can be extended to any complex spin-based circuit.
\end{abstract}
\maketitle

\section{Introduction}
Nanomagnets offer the ultimate thermodynamic limits of computation, as expressed by Landauer's principle~\cite{mayergoyz2009nonlinear}. Therefore, spintronics technology that uses electron spin for information processing and communication presents a favorable option to implement future low-power devices. Most compelling proposals of spintronic devices rely on either the spin-transfer torque (STT) or the dipolar coupling effects between nanomagnets to accomplish switching of the logic device~\cite{edselc.2-52.0-3384774341720050101, 570753420011116, 1308.274520130812, chang2014design, kanimodel, kani2015analysis}. The analysis presented in this paper considers nanomagnet reversal through STT effect~\cite{behin2011switching, srinivasan2011all}. 

A major challenge for spintronics logic, in general, is the inherent stochasticity in nanomagnet dynamics resulting from (i) the presence of thermal noise and (ii) the existence of two basins of attraction that make the dynamic evolution of magnetization extremely sensitive to its initial state~\cite{bertotti2003critical, bertotti2013probabilistic, pinna2013spin, liu2010ultrafast, sankey2008measurement}. There is a two-fold complexity associated with spintronic circuit design due to the stochastic nanomagnet behavior as discussed in our prior work~\cite{kani2014pipeline}. First, the circuit delay exhibits large variability that can increase the effective-delay of the circuit depending on the desired error tolerance. Second, for interconnected logic networks, the delay distributions of individual nanomagnets combine non-linearly increasing the circuit delay complexity significantly~\cite{liu2014dynamics, yao2012magnetic}.

Due to the complex nature of the equation governing the behavior of the nanomagnet body, complete analytical descriptions for the delay distributions of a nanomagnet are unavailable. Previous work has suggested that the effect of thermal noise may be approximated by knowing the initial angle of the nanomagnet and neglecting the thermal noise during reversal \cite{Liuthesis}. While these models do provide significant insight into the nature of nanomagnet reversal, numerical simulations presented in this paper show that these analytical distributions are only accurate for rapid reversal times ($<$ 200 ps). Therefore, for the case when the nanomagnet is under influence of spin currents comparable in magnitude to the critical spin current for reversal, new types of distribution functions to describe the magnetization dynamics must be sought. In this paper, we demonstrate the applicability of the Fr\'{e}chet distribution that contains double exponentials to more accurately capture the evolution of magnetization over a very broad range of time scales. The Fr\'{e}chet distribution is also compatible with the results obtained in~\cite{de2011generalized, gilli2006application} by analytically solving the Landau-Lifshitz-Gilbert (LLG) equation with specific boundary conditions.

Once the nanomagnet delay is characterized using the Fr\'{e}chet distribution function, we evaluate the circuit-level performance of series- and parallel-connected network of nanomagnets. We show that the delay variation is larger when nanomagnets are evaluated in parallel, while a series-connected network yields a much higher average circuit delay. The increase in average delay for series-connected nanomagnets is far greater than any increase in delay one may have to tolerate due to increased delay variation. Therefore, for delay-critical paths, we propose to design spintronic circuits with a larger fan-out and lower logic depth. 

\begin{figure}[tbp]
   \centering
   \includegraphics[width=3.5in]{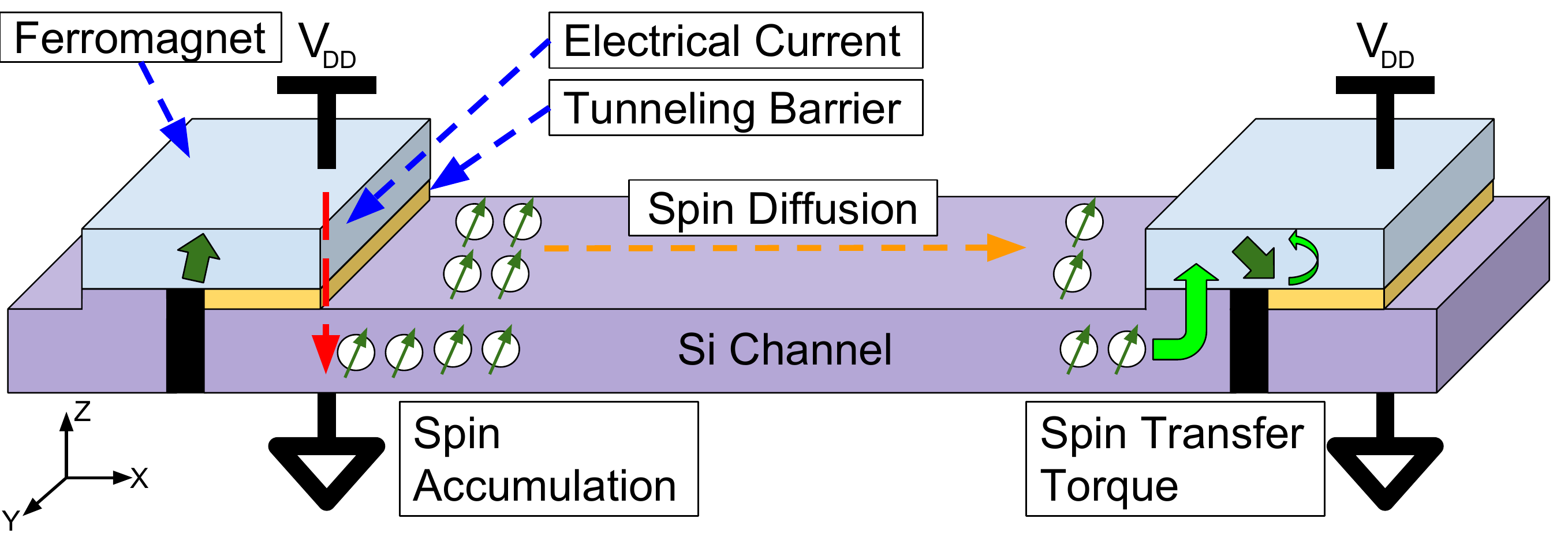} 
   \caption{Schematic of repeater operation for a ASL circuit. An electrical current is driven through the driving nanomagnet (left) and a net spin current is accumulated below the nanomagnet. This spin current then diffuses through the channel and is then delivered to the receiving nanomagnet. For this paper, it is assumed that the channel is long/ressistive enough so that the leakage current can be omitted.}
   \label{fig:ASL_op}
\end{figure}

We use the PDF of the nanomagnet delay to analyze the performance of an ASL-AND gate. 
In an ASL device, an electric current flowing through a transmitting nanomagnet is used to generate a spin current in a non-magnetic channel. Upon reaching the receiving nanomagnet, the spin current imparts a torque to the nanomagnet. The operation of the ASL device is pictorially represented in Fig.~\ref{fig:ASL_op}. Majority Boolean logic gates, such as AND/OR, are implemented by connecting multiple input nanomagnets to a receiving nanomagnet, where a sum of the input spin currents determines the functionality of the logic gate~\cite{kani2014pipeline, edselc.2-52.0-7996091784620110801}.  
 
Due to the summing nature of ASL, the magnitude of the spin current delivered to the output nanomagnet will vary depending on the input values and will alter the delay distributions of the output nanomagnets. The delay distributions presented in this work incorporate the impact of input signal pattern on the overall circuit delay.

\vspace{0.35cm}

\section{Nanomagnet Model}
The ASL device consists of two main components: nanomagnets to store data and transmit logical values and spin channels to allow the transmission of spin current. In this section, we discuss the physical model and the simulation parameters of the nanomagnet. The model for the channel in the ASL device and considerations of dipolar coupling field are presented in the appendices.

\subsection{Nanomagnet Model}
The phenomenological equation describing the physics of a nanomagnet under the effects of STT is the LLG equation \cite{1314667720121201, slonczewski1996current, berger1996emission, landau1935theory, gilbert2004phenomenological}, which is stated as
\begin{equation} \label{eq:LLG}
\frac{d\hat{m}}{dt}=-\gamma {{\mu }_{0}}\left( \hat{m}\times {{{\vec{H}}_{eff}}} \right)+\alpha \left( \hat{m}\times \frac{d\hat{m}}{dt} \right)+\frac{{{{\vec{I}}_{s,\bot }}}}{q{{N}_{s}}},\
\end{equation}
where $\hat{m}$ is the unit-vector along the direction of the magnetization, $\gamma$ is the gyromagnetic ratio, $\mu_{0}$ is the free space permeability, $\alpha$ is the Gilbert damping coefficient, $q$ is the electron charge, $M_{s}$ is the saturation magnetization, and ${I}_{s,\bot }$ is the spin current perpendicular to the magnetization. It is important to note that ordinarily in Magnetic-Tunnel-Junction (MTJ) structures, a field-like (F-L) torque must be considered \cite{sankey2008measurement, kubota2008quantitative, roy2012metastable}. However, as mentioned in Appendix A, only metallic channels are considered in this work; therefore, the F-L torque ids negligible and is not considered in this work. For alternate ASL structures where this F-L torque is significant, the delay distribution must be reevaluated. However, the idealized situation considered in this manuscript is a useful introduction in obtaining the delay distributions of complex spintronic circuits. 

An $\alpha=0.01$ is chosen for the simulations in this paper. Vectors denoted with a ($\hat{.}$) are unit vectors and have a constant magnitude of unity,  while vectors denoted with an ($\overrightarrow{}$) have variable magnitudes. $N_{s}$ is the number of spins in the magnet and is given as 
\begin{equation} \label{eq:Ns}
{{{N}_{s}}=\frac{2{{M}_{s}}V}{\gamma \hbar}}, \
\end{equation}
where $V$ is the volume of the magnet. In (\ref{eq:LLG}), $\vec{H_{eff}}$ is the effective field on the nanomagnet and is given as
\begin{equation} 
{{\vec{H}}_{eff}}={{\vec{H}}_{K}}+{{\vec{H}}_{D}}+{{\vec{H}}_{T}}. 
\label{eq:Field_sum}
\end{equation}
The uniaxial anisotropy field, $\vec{H}_K$ is given from the Stoner-Wahlforth model as \cite{tannous2008stoner}
\begin{equation} \label{eq:Stoner}
{{\vec{H}}_{K}}=\left( \frac{2{{K}_{u}}}{{{\mu }_{0}}{{M}_{s}}}{{m}_{z}} \right)\hat{z},\
\end{equation}
where $K_{u}$ is the uniaxial anisotropy energy density. 
The focus of this work is on thin-film PMA nanomagnets; hence, the anistropy field is assumed to be in the $\hat{z}$ direction. $\vec{H}_{D}$ is the field due to the shape anisotropy (demagnetization) and is given as
\begin{equation} \label{eq:Field_demag}
{{\vec{H}}_{D}}=-{{M}_{s}} \left\langle{{N}_{x}}{{m}_{x}}, {{N}_{y}}{{m}_{y}}, {{N}_{z}}{{m}_{z}}\right\rangle,\
\end{equation}
where $N_{x}$, $N_{y}$, and $N_{z}$ are the demagnetization factors determined by the shape of the nanomagnet \cite{beleggia2006equivalent}. For a thin-film nanomagnet, such as the one assumed in this paper, the demagnetization field is largely in the $\hat{z}$ direction; therefore, we assume $N_{x} = N_{y} = 0$, and $N_{z} = 1$. 

The thermal noise manifests itself as fluctuations in the internal anisotropy field and is added to the internal field of the magnet through the term $\vec{H}_T$ in (\ref{eq:Field_sum})~\cite{brown1963thermal}.
According to the theoretical formulation presented in \cite{1060329}, the thermal field can be defined as an isotropic vector process \cite{gardiner1985handbook}. 
The thermal field can be modeled as a three-dimensional Wiener process \cite{6818391}:
\begin{equation} \label{eq:Field_Thermal}
{{\overset{\lower0.5em\hbox{$\smash{\scriptscriptstyle\rightharpoonup}$}}{H}}_{T}}=\sqrt{\frac{2\alpha{{k}_{B}}T}{\mu_{0}^{2}\gamma{{M}_{S}}V}}\left(\frac{\partial{{W}_{X}}}{\partial t}\hat{x}+\frac{\partial{{W}_{Y}}}{\partial t}\hat{y}+\frac{\partial{{W}_{Z}}}{\partial t}\hat{z}\right),\
\end{equation}
This result is equivalent to the expressions derived using the Fokker-Planck analysis \cite{mayergoyz2009nonlinear}.

The LLG equations are numerically simulated using the Heun numerical method implemented in custom CUDA simulations \cite{6818391, januszewski2010accelerating}. The accuracy of the simulator was verified by comparing results of the simulations against known analytical solutions for the sLLG shown in Section III. We define the nanomagnet reversal delay as the difference in time it takes for the nanomagnet to cross the $\hat{x}-\hat{y}$ plane from the time the spin current was initially applied to the nanomagnet.  
For all simulations in this paper, nanomagnet dimensions are chosen as 100 nm $\times$ 100 nm $\times$ 4 nm, $M_s$ = $3 \times 10^{5}$ $\frac{A}{m}$~\cite{toney2003high}, $K_u$ = $6 \times 10^{5}$ $\frac{J}{m^{3}}$~\cite{toney2003high}, $\alpha$ = 0.01, $I_c$ = 1.32 mA, and input spin current $I_{op}$ = 1.5 mA. Simulations are carried at a lattice temperature of 300 K. 

This work considers idealized nanomagnet and channel structures. Fabrication imperfections such as edge roughness or pinning sites may alter the results. However, analysis under idealized conditions is a useful first step in evaluating complex nanomagnet systems and allows us to benchmark the upper limit of device performance.

In addition, a single-domain approximation is used to model the reversal mechanics of the nanomagnet. Since a magnetic body typically consists of multiple domains that interact through an exchange field, the magnetization evolution can be incoherent \cite{tannous2008stoner, kani2015analysis}. While the effect of incoherency is omitted from the delay distribution analysis, the models presented in this work have been fitted to experimental data and can still be applied to real-world setups assuming some fitting parameters \cite{liu2014dynamics}.   

\section{Single Nanomagnet Delay Distribution Models}
The physics of PMA nanomagnet reversal has been exhaustively studied in \cite{pinna2013thermally, pinna2013spin}. Assuming a perfectly square nanomagnet, the energy landscape of the nanomagnet is greatly simplified allowing analytical solutions of the LLG equation. The implicit analytical equation for the dynamics of the nanomagnet is given as~\cite{liu2014dynamics}
\begin{eqnarray} 
\left ( i - 1 \right )\frac{\tau}{\tau_{D}} = \ln\left ( \frac{\tan\left ( \frac{\phi_{\tau}}{2} \right )}
{\tan\left ( \frac{\phi_{0}}{2} \right )} \right ) \nonumber \\
-\frac{1}{i+1}\ln\left ( \frac{\frac{i-1}{i+1}+\frac{\tan^{2}\left ( \phi_{\tau } \right )}{4}}{\frac{i-1}{i+1}+\frac{\tan^{2}\left ( \phi_{0 } \right )}{4}} \right ),
\label{eq:PMA_delay}
\end{eqnarray}  
where $\tau$ is the time it takes for the polar angle of the magnetization to transition from $\phi = \phi_{0}$ to $\phi = \phi_{\tau}$. $\tau_{D}$ is the time scale of the magnetization dynamics and is given as
\begin{equation} \label{eq:PMA_delay_time_const}
\tau_{D} = \left ( \frac{1 + \alpha^{2} }{\alpha \gamma \mu_{0}\left ( H_{k}-M_{s} \right ) } \right ).  \ 
\end{equation}  
In (\ref{eq:PMA_delay}), $i = I/I_{C}$ is the ratio of the spin current entering the nanomagnet and the critical spin current of the nanomagnet. $I_{c}$ is mathematically given as \cite{mangin2006current}
\begin{equation} \label{eq:PMA_delay_crit_current}
I_{C} = \frac{2 e M_{s} V \alpha}{\hbar}\mu_{0} \left ( H_{k} - M_{s} \right ),
\end{equation}  
where $e$ is the elementary charge. 
Eq. (\ref{eq:PMA_delay}) reduces to the expression derived in the seminal work of J.Z. Sun~\cite{JZSUN2000} under the condition that the input spin current of the nanomagnet vastly exceeds its critical current.

\subsection{Analytical PDF for Rapid Reversals} 
Assuming a large energy barrier between the two stable states of the nanomagnet, the probability distribution of the initial angle of the magnetization of the nanomagnet is given as \cite{pinna2013thermally} 
\begin{subequations}
\begin{equation} 
P(\phi) = e^{-\xi \phi^{2}},
\label{eq:init_ang_pdf}    
\end{equation}  
\begin{equation}
\xi = \frac{\mu_{0} M_{s} V H_{k}}{2 k_{B}T}.
\label{eq:init_ang_X}
\end{equation}
\end{subequations}
During fast reversals, it is expected that the thermal noise has little effect on the nanomagnet during its transition. Instead, the thermal noise only sets the initial angle, which affects the reversal delay according to (\ref{eq:PMA_delay}). Using this assumption, three different probability distribution models for nanomagnet switching were derived in~\cite{Liuthesis}. The analytical cumulative distribution functions (CDFs) are given below for completeness:
\begin{subequations}
\begin{equation} 
P = \textup{exp}\left \{ - 4 \xi \left ( \frac{i-1}{2i} \right )^{2\left ( i + 1 \right )} \textup{exp}\left [ -\left ( i-1 \right )\frac{2\tau }{\tau_{D}} \right ] \right \} 
\label{eq:PDF_delay_1} 
\end{equation}
\begin{equation}
P = \textup{exp}\left \{ - 4 \xi \textup{exp}\left [ -\left ( i-1 \right )\frac{2\tau }{\tau_{D}} \right ] \right \} 
\label{eq:PDF_delay_2}
\end{equation}
\begin{equation} 
P = \textup{exp}\left \{ - \frac{\pi^{2} \xi}{4} \textup{exp}\left [ -\left ( i-1 \right )\frac{2\tau }{\tau_{D}} \right ] \right \} 
\label{eq:PDF_delay_3}
\end{equation}
\label{eq:PDF_delay}
\end{subequations}
Eq. (\ref{eq:PDF_delay_1}) corresponds to the derivation of the CDF through (\ref{eq:PMA_delay}) assuming a large energy barrier. The PDFs of the delay can be found by taking the derivative of the CDFs. Note that the CDFs are denoted by ``$P$'' while PDFs are denoted by ``$p$''. Assuming $i >> 1$, (\ref{eq:PDF_delay_1}) can be further simplified to (\ref{eq:PDF_delay_2}). Finally, (\ref{eq:PDF_delay_3}) assumes both the initial and final magnetization angles are small. 

\begin{figure*}[tbhp]
   \centering
   \includegraphics[width = 7in]{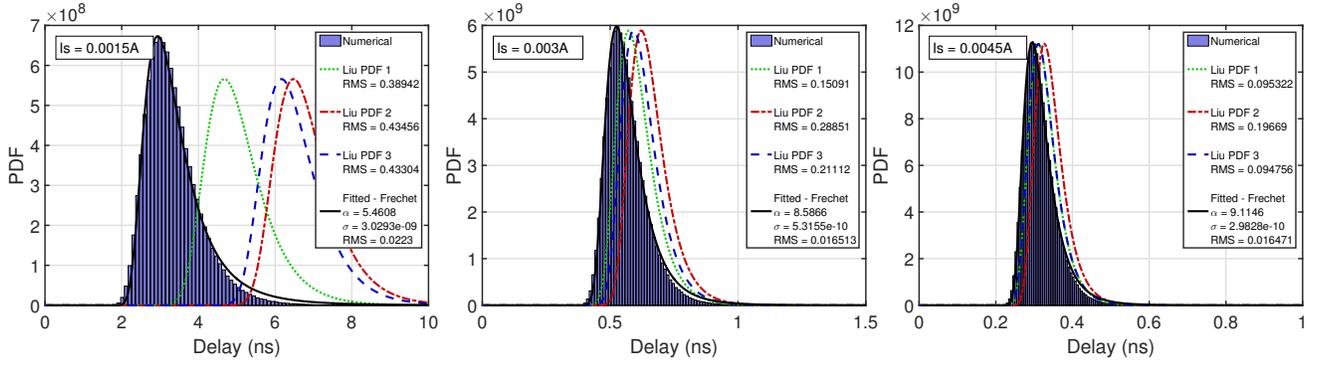} 
   \caption{Switching delay distributions for a 100 nm $\times$ 100 nm $\times$ 4 nm, PMA-type nanomagnet with $M_s$ = $3 \times 10^{5}$ $A/m$ and $K_u$ = $6 \times 10^{5}$ $\frac{J}{m^{3}}$ under the influence of an anti-parallel longitudinal spin current with magnitude as shown in figure labels. PDFs 1, 2, and 3 correspond to equations (\ref{eq:PDF_delay_1}), (\ref{eq:PDF_delay_2}), and (\ref{eq:PDF_delay_3}) respectively. As you can see, PDFs are only accurate when the input spin current is much larger than the critical current. The Fr\'{e}chet distribution is also fitted to each of the data sets and very accurately describes all three delay distributions. Each subplot has $10^{6}$ data points. The normalized root-mean-square (NRMS) values are calculated for each distribution against the numerical data. The maximum of the numerical PDF is used as the normalization factor of the RMS value. }
   \label{fig:Lui_PDF_comp}
\end{figure*}

Figure \ref{fig:Lui_PDF_comp} compares the PDFs of (\ref{eq:PDF_delay_1}-\ref{eq:PDF_delay_3}) to numerical results. In ~\cite{Liuthesis}, it is shown that an $i > 2$ is sufficient to accurately describe the delay variation of the nanomagnet. However, in this work we note that these PDFs only become accurate at much larger values of $i$. This effectively means that the analytical PDFs are accurate only when the reversal time of the nanomagnet is $<$ 100 ps. Under such conditions, the electrical current required at the transmitting nanomagnet would vastly exceed the maximum threshold for electromigration of both the nanomagnet and the non-magnetic metallic channel; therefore, there will be reliability concerns that will reduce the mean-time-to-failure of the ASL device. 
Hence, other PDFs are needed to capture the magnetization dynamics accurately for reversal times on the order of several hundreds of picoseconds or nanoseconds. 

\subsection{Analytical PDF for Near-Threshold Reversals} 
The analytical solution of the LLG becomes formidable for conditions where the input spin current is comparable to the critical spin current of the nanomagnet.
To obtain a PDF that best represents the nature of the delay variation of the nanomagnet, we consider the following situation. 
It is shown in the previous sub-section that as the spin current is increased, the PDF of the delay tends toward a double exponential function. In addition, it is known that if the spin current drops below the critical current, nanomagnet reversal occurs only when the magnetization angle of the nanomagnet becomes large enough such that the sub-critical current can overcome the reduced energy barrier of the nanomagnet\cite{1060329}. This process of nanomagnet reversal, primarily through thermal activation, is known to be a single exponential function. Therefore, we seek a PDF solution to describe the nanomagnet reversal that can be seamlessly adjusted from a single- to a double-exponential function depending on the value of $i$. In this work, we consider the Fr\'{e}chet distribution that meets the above criteria. The Fr\'{e}chet distribution is mathematically given as~\cite{harlow2002applications}
\begin{equation} \label{eq:PDF_Frechet}
p = \frac{\alpha }{s}\left ( \frac{\tau}{s} \right )^{-1-\alpha }e^{\left ( -\frac{\tau}{s} \right )^{-\alpha }},  
\end{equation}
where $s$ is the scale parameter, $\alpha$ is the shape parameter, and $\tau$ is the delay of the nanomagnet. For the purposes of this paper, $\alpha$ and $s$ are treated as fitting parameters~\cite{mann1984statistical}. To prove the suitability of the Fr\'{e}chet distribution, Fig.~\ref{fig:Lui_PDF_comp} shows the best fits of the Fr\'{e}chet distribution to the delay curves of a nanomagnet driven by various spin currents~\cite{johnson2000probability}. The figure clearly shows improved accuracy of the Fr\'{e}chet distribution to capture the numerical simulation data and the applicability of the Fr\'{e}chet distribution to nanomagnet delay under various reversal regions. Fig.~\ref{fig:Lui_PDF_comp} demonstrates that the $\alpha$ parameter increases greatly if $i >> 1$. Since a smaller $\alpha$ suggests a larger left lean, this suggests that reversal distributions for nanomagnets under large spin currents have less left lean.   

For the remainder of this paper, the delay of a nanomagnet under the influence of a specific critical field will be represented by a Fr\'{e}chet distribution that has been fitted to numerical data. 

\subsection{Relationship to Error-Rate} 

Because the delay has been shown to be a random variable, the probability that the nanomagnet delay will exceed some time $t$ will always be nonzero. When designing circuits, this probability can be referred to as the nanomagnets error rate ($er$). Using the Fr\'{e}chet distribution (\ref{eq:PDF_Frechet}) it is possible to derive $t$ as a function of $er$. Mathematically, this relationship is given as

\begin{equation} \label{eq:error_rate_t}
t=\frac{s}{\sqrt[\alpha]{-\ln\left [ 1-er \right ]}},  
\end{equation}
where $\alpha$ and $s$ are defined previously in (\ref{eq:PDF_Frechet}).

\section{Combination of Reversal Delay Distributions}
In any complex circuit, devices can be arranged in two ways. They can be operated in parallel, where their outputs arrive at the same time. Alternatively, devices can be cascaded in series, where the output of one is fed into the input of of another. To analyze the total delay of a circuit, both these cases need to be studied carefully. 

\subsection{Devices in Parallel}
\begin{figure}[h!]
   \centering
   \includegraphics[width=3.5in]{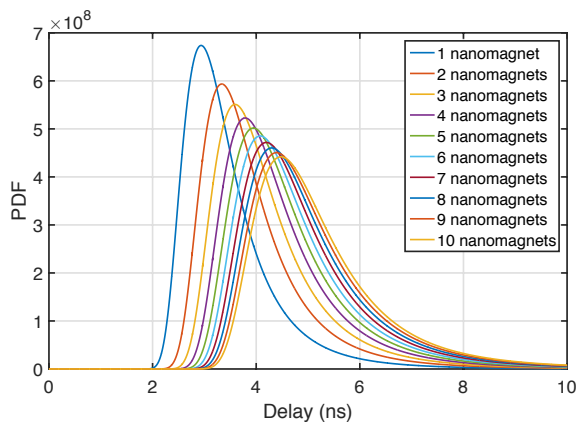} 
   \caption{Demonstrates the reversal delay of a circuit composed of multiple nanomagnets arrange in parallel. A 100 nm $\times$ 100 nm $\times$ 4 nm, PMA-type nanomagnet with $M_s$ = $3 \times 10^{5}$ $A/m$ and $K_u$ = $6 \times 10^{5}$ $\frac{J}{m^{3}}$ and is under the influence of a $1.5 mA$ anti-parallel spin current. The pdf of a single nanomagnet is estimated by fitting a Fr\'{e}chet distribution to numerical data. Multi-nanomagnet results are calculated numerically.}
   \label{fig:multi_par}
\end{figure}

For the case of a circuit with multiple devices in parallel, the output delay (denoted by random variable $Y$) of the circuit is the maximum of the output delay of the devices given as
\begin{equation} \label{eq:Par_PDF_1}
Y = \textup{max}\left \{ X_{1}, X_{2}, X_{3}, ... X_{n} \right \}.
\end{equation}  
where $X_{i}$ is a random variable representing the delay of a single nanomagnet. Hence, the CDF of the parallel device circuit becomes
\begin{equation} 
P\left ( Y \leq x \right ) = P\left ( X_{1} \leq x,..., X_{n} \leq x \right ) \\ = \prod_{i=1}^{n}P(X_{i} \leq x) = \left ( P\left ( x \right ) \right )^{n},
\label{eq:Par_PDF_2}
\end{equation}  
where $P$ corresponds to the CDF of the particular device. Knowing this, the PDF of a circuit with multiple devices in parallel can be found using (\ref{eq:Par_PDF_2}) and (\ref{eq:PDF_Frechet}) and is plotted in Fig.~\ref{fig:multi_par}. Assuming a Fr\'{e}chet distribution, the PDF of multiple nanomagnets in parallel is given analytically as
\begin{equation} 
p = n \frac{\alpha }{s} \left ( \frac{x}{s} \right )^{-1-\alpha } e^{n\left ( -\frac{x}{s} \right )^{-\alpha }},
\label{eq:Par_PDF_Frechet}
\end{equation}  
where $n$ is the number of devices in parallel.  

\subsection{Devices in Series}
\begin{figure}[h!]
   \centering
   \includegraphics[width=3.5in]{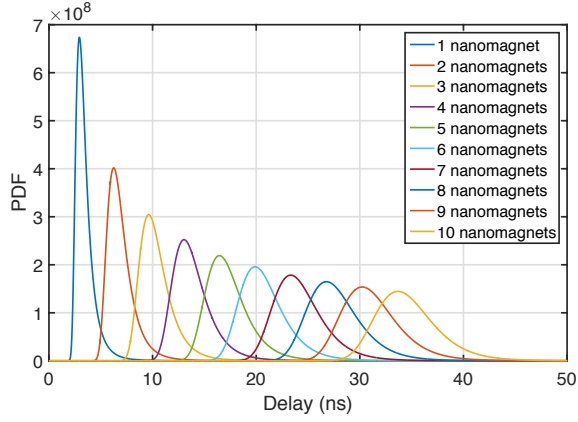} 
   \caption{Demonstrates the reversal delay of a circuit composed of multiple nanomagnets arrange in series. A 100 nm $\times$ 100 nm $\times$ 4 nm, PMA-type nanomagnet with $M_s$ = $3 \times 10^{5}$ $A/m$ and $K_u$ = $6 \times 10^{5}$ $\frac{J}{m^{3}}$ and is under the influence of a $1.5 mA$ anti-parallel spin current. The pdf of a single nanomagnet is estimated by fitting a Fr\'{e}chet distribution to numerical data. Multi-nanomagnet results are calculated numerically.}
   \label{fig:multi_ser}
\end{figure}

For the case of devices connected in series, the output delay (Y) of the circuit is the addition of the device delays connected in series and is given as
\begin{equation} 
Y = X_{1} + X_{2}. 
\label{eq:Ser_PDF_1}
\end{equation}  
The PDF of Y is given as
\begin{equation} 
p_{Y}\left ( y \right ) = \int_{0}^{y} p_{X_{2}}\left ( y-x \right )p_{X_{1}}\left ( x \right )dx 
\label{eq:Ser_PDF_2} 
\end{equation}  
Figure \ref{fig:multi_ser} shows the PDFs of multiple nanomagnet devices arranged in series. This situation corresponds to a repeater-chain circuit. As the number of devices increases, the distribution of the delay gets more symmetrical and normal, following the central limit theorem~\cite{rice2006mathematical}.
 
\subsection{Comparison of Device Arrangments} 
\begin{figure}[h!]
   \centering
   \includegraphics[width=3.5in]{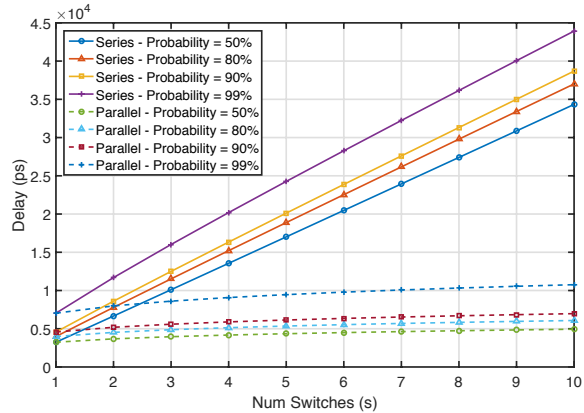} 
   \caption{The reversal delay of a circuit comprising multiple nanomagnets arranged in series and parallel versus the number of switches. While arranging devices in parallel slightly increases delay of the circuit, this increase is minimal compared to the average delay of the nanomagnet. Multiple error rates are considered.}
   \label{fig:multi_comp}
\end{figure}

Often when developing spin-based circuits, it is possible to achieve similar functionalities using many devices driven in parallel, or many devices cascaded off eachother. A prime example of this is the many variations of VLSI adder designs. Since devices arranged in parallel increase the nanomagnet variation, it can be argued that highly parallel circuits may have longer delays than serialized circuits given a particular $er$. However, Fig. \ref{fig:multi_comp} demonstartes that this not the case. While the number of devices in parallel does increase the circuit delay, this increase is still dwarfed by the increase in average delay caused by arranging the nanomagnet in series.

\section{AND-gate Analysis}
The previous sections introduced the key concepts required to analyze the delay variation of a complex logic circuit. For this paper, a four-input ASL-AND circuit is designed and analyzed. An example of a four-input AND (AND4) gate is shown in the inset of Fig.~\ref{fig:pdf_AND4_par}. As mentioned earlier, the summing nature of the inputs in ASL logic naturally create a majority logic. To create an AND gate from this majority logic, the input of the logic nanomagnet must be weighted such that the input to the nanomagnet is only positive when all the inputs are positive. In the case of the ASL-AND circuit shown, a fixed magnet generating a constant $-3 \times I_{s}$ is added such that the output nanomagnet will only receive a $+\hat{z}$-orientated spin current if all the nanomagnet are oriented along $+\hat{z}$. The polarity of this bias can be changed to switch between AND and OR logic. It is assumed that the input nanomagnets have been reversed at time $\tau = 0$ and the voltage supplies at these nanomagnets are turned on at this time \cite{calayir2014static}. 

\subsection{Effect of Input Pattern on Nanomagnet Delay}
An important aspect of current-based computation is the fact that the input current to the OUT nanomagnet will not be constant. In fact, depending on the logical combination of the input devices, the current being fed into the OUT nanomagnet will vary.
Assuming the inputs of the nanomagnet are equi-probable, the likelihood that the OUT nanomagnet is reversed by a spin current of a particular magnitude is given by Table~\ref{tab:in_prob_cur_2}. The probabilities of each of these magnitudes follows an ordering, which can be described by Pascal's Triangle. 

\subsection{ASL-AND Delay-PDF}
\begin{figure}[tbhp]
   \centering
\includegraphics[width=3.5in]{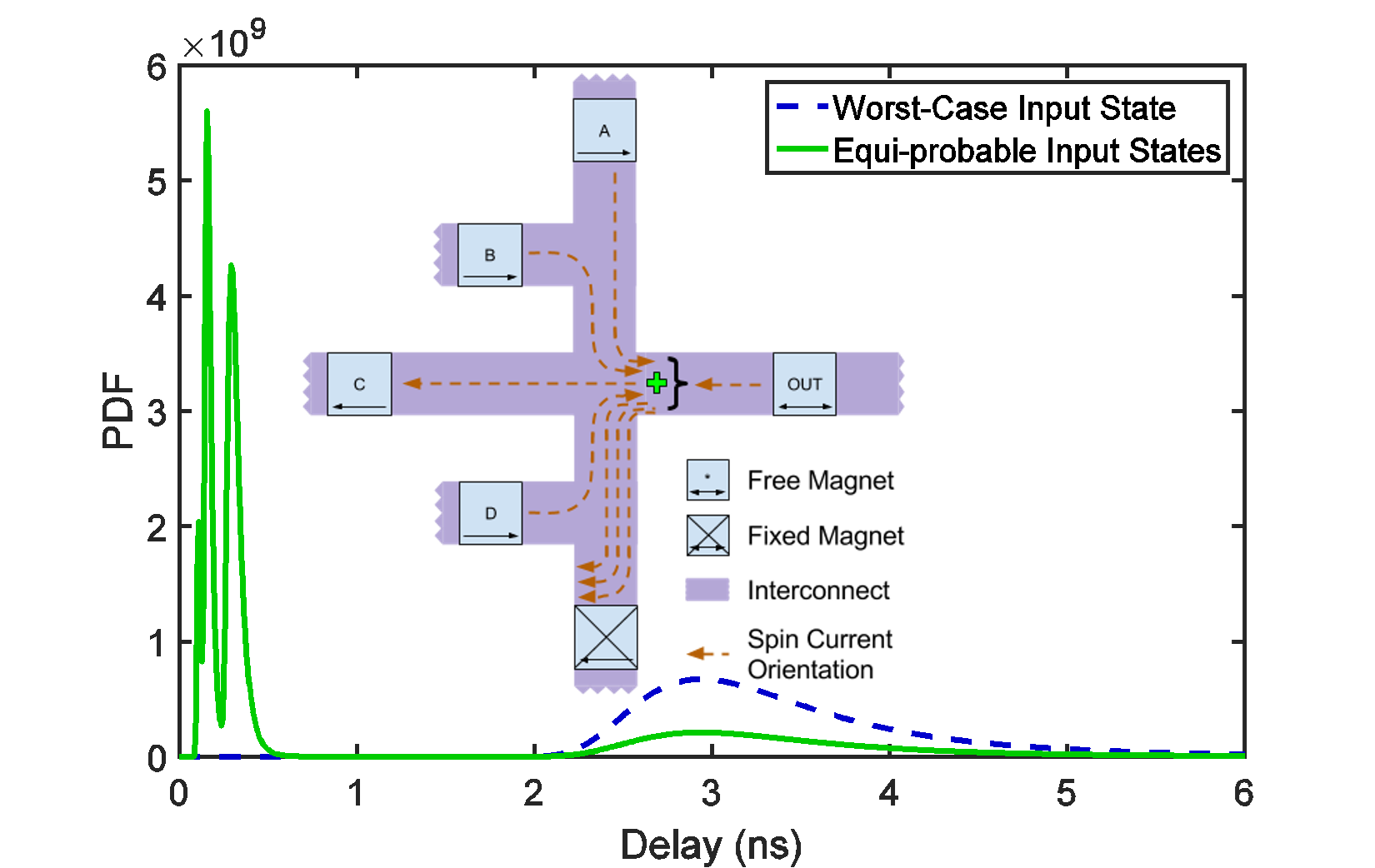}  
   \caption{PDFs of parallelized AND4 circuit assuming several different types of delay analyses. Baseline PDF simply considers the delay of the AND-gate assuming worst-case input conditions (OUT nanomagnet being driven by $I_{s}$. Input-probability aware distribution analyzes the OUT nanomagnet reversal considering the variable input spin current magnitudes considering different input combination. Inset: Schematic of parallelized ASL AND4 gate. Blue square represent thin-film nanomagnets while purple bars represent channels. Orange arrows represent spin orientation.}
   \label{fig:pdf_AND4_par}
\end{figure}

At worst, the nanomagnet will be driven by a current magnitude $ = I_{s}$. This corresponds to the case where all the inputs, except one, are oriented along the $+\hat{z}$ direction, and only one of the input nanomagnets is oriented along the $-\hat{z}$ direction. This worst-case scenario is shown in Fig.~\ref{fig:pdf_AND4_par}. However, as mentioned previously, depending on the input pattern, the driving spin current to OUT is likely larger than $I_{s}$. For larger spin current magnitudes, the nanomagnet is expected to reverse over much shorter timescales. By calculating the PDFs of nanomagnet reversal at each of the different spin current magnitudes (by fitting the Fr\'{e}chet distribution to  numerical results), the PDFs can be combined to find the input-aware PDF of the circuit delay as shown in Fig.~\ref{fig:pdf_AND4_par}. This new input-aware PDF has several peaks corresponding to each of the possible spin current magnitudes during the operation of the circuit. As the number of inputs increases, the number of peaks will also increase, but the area under each of the peaks will decrease since the area of the entire PDF must remain equal to unity. Assuming a very small error-rate, one is mainly concerned with the right-most peak that is associated with nanomagnet reversal under the minimum operating current. This suggests that for a given delay, the circuit reliability is improved as the number of inputs increases. In other words, as the circuit becomes more complex, it tends to operate more reliably given a certain delay.

\begin{table}[]
\centering
\caption{Assuming all logical input combinations are equally possible, table demonstrates the probability that a nanomagnet is reversed with a particular spin current magnitude. Spin current magnitude is normalized against minimum reversal current.}
\label{tab:in_prob_cur_2}
\begin{tabular}{|l | l|}
\hline
Current Magnitude & Probability    \\ \hline
+1                & $\frac{1}{16}$ \\
-1                & $\frac{4}{16}$ \\
-3                & $\frac{6}{16}$ \\
-5                & $\frac{4}{16}$ \\
-7                & $\frac{1}{16}$ \\ \hline
\end{tabular}
\end{table}

\vspace{0.35cm}

\section{Conclusion}
In this paper, analytical and numerical methods have been developed to analyze the delay variation of a nanomagnet-based circuit by using the probability distribution functions that describe the stochastic dynamics of the nanomagnet. 
Firstly, the reversal delay of a single PMA nanomagnet under the effects of thermally-induced stochasticity is extensively analyzed. It is shown that the previously published models are mainly accurate in the limiting case of large input spin current or short reversal timescales of the nanomagnet.
Here, we study the reversal of nanomagnets that are operating near critical conditions that allow reversal over several hundreds of picoseconds or nanosecond timescales. We use the Fr\'{e}chet distribution function to model the nanomagnet reversal over a broad range of input spin currents and reversal times. The Fr\'{e}chet distribution is fitted to numerical simulation data obtained from solving the stochastic LLG equation. Note, the delay distributions analyzed here are for a single-domain nanomagnet. While \cite{liu2014dynamics} found that the probability density functions of (\ref{eq:PDF_delay}) can be fitted to experimental reversal data, further work is needed to understand how magnetization incoherency will affect the delay distribution curves. 

Knowing the delay of a single nanomagnet, it is possible to analyze nanomagnet circuits where nanomagnets are evaluated in either parallel or series configurations. It is shown that a circuit with nanomagnets evaluated in parallel has a larger variation but a smaller average delay than a circuit with nanomagnets in series. Finally, these concepts are applied to the evaluation of a ASL-AND gate. It is shown that having more inputs to an AND circuit is beneficial since the fraction of the time the nanomagnet is operating under the minimum input spin current is reduced. 

The work in this paper is an initial step towards building and evaluating much larger nanomagnetic circuits. Further work is need to evaluate the delay variation of circuits implemented in alternate magnetic technologies and evaluating the performance metrics of complex spintronic logic circuits. 

\section*{Appendix-A: Channel model in ASL device}
\begin{figure}[h!]
\centering
\includegraphics[width=3.5in]{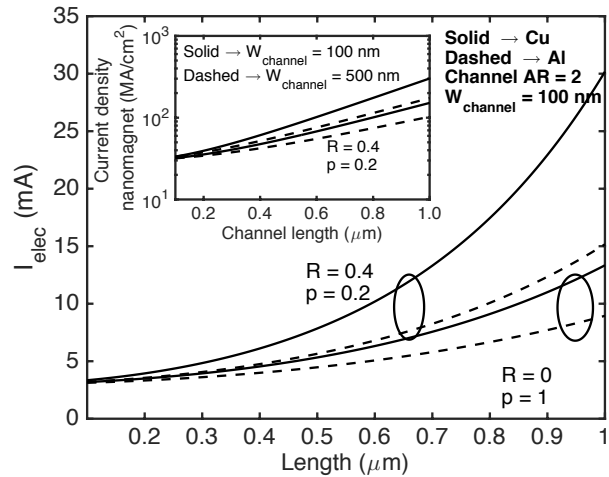} 
\caption{Required electrical current in the ASL device to achieve 1.5 mA of spin current at the receiver as a functio of the channel length. Different values of size-effect parameters are considered. The inset plot shows the corresponding electrical current density through the nanomagnet. The spin polarization of the nanomagnet is assumed to be 0.5. Other simulation parameters are noted in the figure.}
\label{fig:current_combined_nm}
\end{figure}

The amount of electric current, $I_{elec}$, pumped into the transmitter that reaches the receiver is quantified through spin injection and transport efficiency (SITE). Here, we use the mathematical models for SITE derived in \cite{rakheja2013impact, rakheja2013roles} to obtain the amount of $I_{elec}$ required to achieve a specific amount of spin current at the receiver. The models take into account size effects in ultra-scaled metallic channels.
As shown in Fig.~\ref{fig:current_combined_nm}, for a channel length of 500 nm and in the absence of grain-boundary ($R = 0$) and sidewall scatterings ($p = 1$), the amount of electrical current to obtain 1.5 mA of spin current at the receiver nanomagnet is 5.3 mA and 4.5 mA for copper and aluminum channels, respectively. The required electrical current increases in the presence of realistic size effects. The inset plot of Fig.~\ref{fig:current_combined_nm} shows the electrical current density through the nanomagnet as a function of channel length for different values of channel width. While increasing the width of the channel reduces the electrical current density through the nanomagnet and improves reliability of the ASL device, it also increases the overall device footprint and will limit the device scalability.

The delay associated with spin diffusion through the channel is given as 
\begin{eqnarray}
t_{diff} = \frac{L_{int}^2}{2D_s},
\end{eqnarray}
where $D_s$ is the diffusion coefficient of electrons in the channel. Using $D_s$ = 126 $cm^2/Vs$ and 80 $cm^2/Vs$ for Cu and Al, respectively, the diffusion delay through a 500-nm long spin channel is only about 20 ps~\cite{rakheja2013impact}. This delay is more than an order of magnitude lower than the nanomagnet switching delay and will not be considered in this work.

\section*{Appendix-B: Dipolar Coupling}
It has been previously argued that in ASL circuits, nanomagnet coupling becomes an issue for only very short interconnects~\cite{scalingallspin}. At channel lengths of several hundreds of nanometer considered in this work, the dipolar field generated by one nanomagnet to the center of the other would be $~0.1\% H_{k}$ \cite{kanimodel, 1883308}. Since the magnitude of the dipolar field is three orders of magnitude smaller than the $\hat{z}$-anistropy of the nanomagnet, the effect of dipolar coupling is omitted in this work.

\vspace{0.35cm}
\begin{acknowledgements}
\vspace{0.35cm}
This project was supported by the Nanoelectronics Research Corporation (NERC), a wholly-owned subsidiary of the Semiconductor Research Corporation (SRC), through the Institute for Nanoelectronics Discovery and Exploration (INDEX). The authors would also like to thank Prof. Andrew Kent and his group at New York University for their help and insight. We would also like to thank Dr. Daniele Pinna from CNRS for useful discussions. 
\end{acknowledgements}


\end{document}